\documentclass[12pt,epsf]{article}
\usepackage{graphicx}
\begin{document}

\def\a{\alpha}
\def\b{\beta}
\def\c{\varepsilon}
\def\d{\delta}
\def\e{\epsilon}
\def\f{\phi}
\def\g{\gamma}
\def\h{\theta}
\def\k{\kappa}
\def\l{\lambda}
\def\m{\mu}
\def\n{\nu}
\def\p{\psi}
\def\q{\partial}
\def\r{\rho}
\def\s{\sigma}
\def\t{\tau}
\def\u{\upsilon}
\def\v{\varphi}
\def\w{\omega}
\def\x{\xi}
\def\y{\eta}
\def\z{\zeta}
\def\D{\Delta}
\def\G{\Gamma}
\def\H{\Theta}
\def\L{\Lambda}
\def\F{\Phi}
\def\P{\Psi}
\def\S{\Sigma}

\def\o{\over}
\newcommand{\gsim}{ \mathop{}_{\textstyle \sim}^{\textstyle >} }
\newcommand{\lsim}{ \mathop{}_{\textstyle \sim}^{\textstyle <} }
\newcommand{\vev}[1]{ \left\langle {#1} \right\rangle }
\newcommand{\bra}[1]{ \langle {#1} | }
\newcommand{\ket}[1]{ | {#1} \rangle }
\newcommand{\EV}{ {\rm eV} }
\newcommand{\KEV}{ {\rm keV} }
\newcommand{\MEV}{ {\rm MeV} }
\newcommand{\GEV}{ {\rm GeV} }
\newcommand{\TEV}{ {\rm TeV} }
\def\diag{\mathop{\rm diag}\nolimits}
\def\Spin{\mathop{\rm Spin}}
\def\SO{\mathop{\rm SO}}
\def\O{\mathop{\rm O}}
\def\SU{\mathop{\rm SU}}
\def\U{\mathop{\rm U}}
\def\Sp{\mathop{\rm Sp}}
\def\SL{\mathop{\rm SL}}
\def\tr{\mathop{\rm tr}}

\def\IJMP{Int.~J.~Mod.~Phys. }
\def\MPL{Mod.~Phys.~Lett. }
\def\NP{Nucl.~Phys. }
\def\PL{Phys.~Lett. }
\def\PR{Phys.~Rev. }
\def\PRL{Phys.~Rev.~Lett. }
\def\PTP{Prog.~Theor.~Phys. }
\def\ZP{Z.~Phys. }

\newcommand{\bear}{\begin{array}}  \newcommand{\eear}{\end{array}}
\newcommand{\bea}{\begin{eqnarray}}  \newcommand{\eea}{\end{eqnarray}}
\newcommand{\beq}{\begin{equation}}  \newcommand{\eeq}{\end{equation}}
\newcommand{\bef}{\begin{figure}}  \newcommand{\eef}{\end{figure}}
\newcommand{\bec}{\begin{center}}  \newcommand{\eec}{\end{center}}
\newcommand{\non}{\nonumber}  \newcommand{\eqn}[1]{\beq {#1}\eeq}
\newcommand{\la}{\left\langle} \newcommand{\ra}{\right\rangle}

\def\SEC#1{Sec.~\ref{#1}}
\def\FIG#1{Fig.~\ref{#1}}
\def\EQ#1{Eq.~(\ref{#1})}
\def\EQS#1{Eqs.~(\ref{#1})}
\def\lrf#1#2{ \left(\frac{#1}{#2}\right)}
\def\lrfp#1#2#3{ \left(\frac{#1}{#2}\right)^{#3}}
\def\GEV#1{10^{#1}{\rm\,GeV}}
\def\MEV#1{10^{#1}{\rm\,MeV}}
\def\KEV#1{10^{#1}{\rm\,keV}}


\baselineskip 0.7cm

\begin{titlepage}

\begin{flushright}
IPMU09-0057 \\
UT-09-12
\end{flushright}

\vskip 1.35cm
\begin{center}
{\large \bf
Conformal Supersymmetry Breaking \\ in Vector-like Gauge Theories
}
\vskip 1.2cm
Izawa K.-I.$^{1,2}$, Fuminobu Takahashi$^{2}$, T. T. Yanagida$^{2,3}$, and Kazuya Yonekura$^{3}$
\vskip 0.4cm
{\it  
${}^1$Yukawa Institute for Theoretical Physics, Kyoto University,\\
Kyoto 606-8502, Japan\\
${}^2$ Institute for the Physics and Mathematics of the Universe,
  University of Tokyo,\\ Chiba 277-8568, Japan\\
${}^3$
Department of Physics, University of Tokyo,\\
    Tokyo 113-0033, Japan}

\vskip 1.5cm

\abstract{A new class of models of dynamical supersymmetry breaking is proposed.
The models are based on $SU(N_C)$ gauge theories with $N_F(<N_C)$ flavors of quarks and singlets.
Dynamically generated superpotential exhibits runaway behavior. By embedding the models into 
conformal field theories at high energies, the runaway potential is stabilized
by strong quantum corrections to the K\"ahler potential. The quantum corrections are large but nevertheless
can be controlled due to superconformal symmetry of the theories.
 
}
\end{center}
\end{titlepage}

\setcounter{page}{2}

\section{Introduction}

Vector-like models of supersymmetry (SUSY) breaking were
sought from the beginning of investigation for dynamical
SUSY breaking~\cite{Witten:1981nf,Witten:1982df,Affleck:1983mk}. However, 
arguments based on the Witten index~\cite{Witten:1982df} suggested that the construction of such a model may be difficult.   
Only very limited models with quantum deformed moduli have been known
so far to provide dynamics for vector-like SUSY breaking~\cite{Izawa:1996pk} (see also~\cite{Dimopoulos:1997ww}).
Even if one takes into account metastable SUSY-breaking vacua present in massive vector-like models~\cite{Intriligator:2006dd},
all the known models of vector-like SUSY breaking were based on gauge theories with flavor numbers larger than or equal to that in the deformed moduli case.
For example, in an $SU(N_C)$ gauge theory with $N_F$ flavors of quarks, 
$N_F \geq N_C$ in such SUSY breaking models.

In a SUSY $SU(N_C)$ gauge theory with $N_F < N_C$ flavors of massless quarks $Q$ and antiquarks
$\tilde{Q}$, the quantum superpotential implies runaway behavior of vacua. 
Because of this runaway behavior, this model was not used to construct a vector-like SUSY breaking model.
However, it has been, recently, shown that the runaway potential is stabilized by quantum corrections 
to the K\"ahler potential in a simple extension of the original SUSY gauge theory~\cite{Izawa:2009mj}. 
The stabilization of the potential leads to a dynamical SUSY breaking.

The model is based on an introduction of a singlet field $S$ with a tree level 
superpotential $W=\l SQ\tilde{Q}$. We see again a runaway potential for $S$ in the large $|S|$ region,
provided the minimal K\"ahler potential for the $S$. However, we should take into account quantum corrections
to the K\"ahler potential for the $S$ to examine the dynamics of the $S$ field, since
the full potential for $S$ is given by both contributions from
the superpotential and the K\"ahler potential. The K\"ahler potential for the singlet $S$ is given by 
the anomalous dimension $\g_S$ of the $S$, 
which is determined by the Yukawa coupling $\l$ at the 1-loop level. 
We find that the potential is an 
increasing function of $|S|$ if ${\tilde \g}_S(|S|)/2>1-N_F/N_C$ is
satisfied ($\tilde{\g_S}$ will be defined later).

However, because ${\tilde \g}_S$ is loop suppressed, it is necessary to tune $N_F$ and $N_C$ such that $1-N_F/N_C \ll 1$ for perturbative computation to be reliable.
Furthermore, in Ref.~\cite{Izawa:2009mj}, the asymptotic non-free nature of  the Yukawa coupling $\l$ is 
utilized to stabilize the potential, which implies that
$\l$ eventually hits a so-called Landau pole at some high energy scale. Then, the theory becomes ill 
defined in the high energy regime.

In this paper, we embed the above SUSY theory into a conformal one. 
Because of (approximate) superconformal symmetry, 
the K\"ahler potential can be well controlled even in a strong coupling regime. Furthermore, there is 
no Landau pole problem since all couplings can be assumed to take the fixed point values.

Our discussion in this paper is mainly concerned with the behavior of the potential at the large $|S|$ regime.
We will see that the potential is an increasing function of $|S|$, so that the potential minima are 
at the small $|S|$ regime. Because couplings are strong there, a detailed investigation of the minima is difficult.
Nevertheless we will argue in the last section that SUSY is certainly broken, and 
our theories are indeed dynamical SUSY breaking models.

\section{Embedding to conformal field theory}

We consider a SUSY $SU(N_C)$ gauge theory with $N_F$ flavors of quarks $Q^i$ and anti-quarks $\tilde{Q}_i$ ($i=1,\cdots,N_F$)
which belong to the fundamental and anti-fundamental representations of
the $SU(N_C)$, respectively \cite{Intriligator:1995au}.
We also introduce 
a gauge singlet chiral superfield $S$.
To promote the model to a superconformal theory,
we introduce additional $N'_F$ flavors of quarks $P^a$ and $\tilde{P}_a$ ($a=1,\cdots,N'_F$) 
which also belong to the fundamental and anti-fundamental 
representations of the $SU(N_C)$, respectively. We adopt a tree level superpotential 
\beq
W=\l SQ^i\tilde{Q}_i+mP^a\tilde{P}_a.
\eeq
When the mass $m$ vanishes, this theory has an infrared conformal fixed point if $N_C,~N_F$, and $N'_F$
satisfy $3N_C/2<N_F+N'_F<3N_C$, as shown in Ref.~\cite{Ibe:2005pj}. (This is an extension of the result of Ref~\cite{Seiberg:1994pq}.
We show the existence of the infrared fixed point perturbatively in \ref{app-A}, and prove the relation $3N_C/2<N_F+N'_F$
by using general properties of conformal field theories later  .)

In a region where the vacuum expectation value (vev) of $S$ is large,
the $Q$ and $\tilde{Q}$ are massive, which we can integrate out. 
Futhermore, in a low-energy regime, the $P$ and $\tilde{P}$ can also be integrated out, and
the low-energy dynamical scale of the $SU(N_C)$ gauge theory is given by
\beq
\L_L^{3N_C}=m^{N'_F}(\l S)^{N_F} \L^{3N_C-N_F-N'_F},
\eeq
where $\L$ is the dynamical scale of the high-energy theory. Then the
low-energy effective superpotential of $S$ is given by
\beq
W_{\rm eff}=N_C\L_L^3=N_C[m^{N'_F}\L^{(3N_C-N_F-N'_F)}(\l S)^{N_F}]^{1 \o N_C}.
\label{eq:effsuper}
\eeq

We take the number of flavors of $Q,~\tilde{Q}$ to be $N_F<N_C$, and in this case the superpotential (\ref{eq:effsuper}) 
is of the runaway type. As discussed in Ref.~\cite{Izawa:2009mj},
we must know the effective K\"ahler potential to determine whether the
total potential (as opposed to superpotential) is a runaway one.
In Ref.~\cite{Izawa:2009mj}, a weak coupling analysis is done to compute quantum corrections to the effective K\"ahler potential.
We do not assume the weak couplings here, and instead, we use a
superconformal symmetry to determine the K\"ahler potential even in a
strongly coupling theory in the following.

\section{Low-energy effective K\"ahler potential}\label{sec:effkahler}

Now let us consider the low-energy effective K\"ahler potential of the singlet $S$ when massive quarks $Q,~\tilde{Q}$ and $P,~\tilde{P}$ are integrated out. 
We consider in the region where the vev of $S$ is such that the mass of $Q,~\tilde{Q}$ is much larger than the mass of $P,~\tilde{P}$.

We will utilize the superconformal symmetry of the theory which is realized in the limit $m \to 0$. 
For this purpose, let us adopt a mass-independent renormalization scheme, where counter terms of dimensionless couplings do not 
depend on mass parameters. This is possible, because any divergent part of amplitudes can be renormalized by
counter terms which do not depend on mass parameters. Thus we can choose (finite as well as infinite part of) counter terms such that they do not 
depend on mass parameters.

In a mass-independent renormalization scheme, $\b$ functions of dimensionless couplings do not depend on mass parameters.
Thus in our case, even if $m \neq 0$, 
we can tune the gauge and Yukawa couplings so that they are just on the fixed point values of the massless theory. So the coupling constants are really
``constant'' (i.e. renormalization group invariant).

\subsection{K{\" a}hler potential in the massless limit}

After integrating out $Q,~\tilde{Q}$ but not $P,~\tilde{P}$, the singlet field $S$ decouples from the gauge sector, 
and our theory becomes an $SU(N_C)$ gauge theory with 
$N'_F$ flavors of quarks $P,~\tilde{P}$.
For $N_F'<N_C$, the corresponding mesonic degrees of freedom
diverge, that is, $|P {\tilde P}| \rightarrow \infty$ as $m \rightarrow 0$.
On the contrary, for $N_F'>N_C$, they are vanishing
as $|P {\tilde P}| \rightarrow 0$ for $m \rightarrow 0$,
which implies that the effective K{\"a}hler potential of $S$
suffers from no divergent contribution in the massless limit.
We assume the latter is the case, hereafter.

The conformal piece of the effective K{\"a}hler potential is given by
\beq
 \ln { \q^2 K_{\rm eff}^{\rm conf} \o \q S \q S^* }
 = -\int^{|S|}_{\mu=M} {\tilde \g}_S(\mu) \, d (\ln \mu), \label{eq:kahlerrenorm}
\eeq
where the renormalization scheme of Ref.~\cite{Coleman:1973jx} has been used%
\footnote{Even in the presence of non zero $m$, we use the same counterterms as those in the massless case
to maintain mass-independent renormalization. Then, the right hand side of Eq.~(\ref{eq:kahlerrenorm}) differs from the effective
K\"ahler potential $K_{\rm eff}$ by the mass dependent terms $K_{\rm
eff}^{\rm mass}$, i.e. $K_{\rm eff}=K_{\rm eff}^{\rm conf}+K_{\rm eff}^{\rm mass}$.} (see also Ref~\cite{ArkaniHamed:1997ut} for a Wilsonian approach).
Here, $M$ is the renormalization point, and
the ${\tilde \g}_S(\m)$ is given in terms of the anomalous
dimension $\g_{S}$ of $S$ by~\cite{Coleman:1973jx}%
\footnote{
Our definition of $\g$ is $-2$ times that used in Ref.~\cite{Coleman:1973jx}.}
\beq
{\tilde \g}_S(\m)/2=\frac{\g_S(\m)/2}{1+\g_S(\m)/2}.
\eeq

In our theory, ${\tilde \g}_S(\mu) = {\tilde \g}_{S*}$ is constant, which is the fixed-point 
value of the massless theory.
In this case, we obtain
\beq
 K_{\rm eff}^{\rm conf}=(1-{\tilde \g}_{S*}/2)^{-2}
 M^{{\tilde \g}_{S*}}|S|^{2-{\tilde \g}_{S*}} \label{eq:effkahler}
 \label{Kconf}.
\eeq

The anomalous dimension $\g_S$ is the one defined in the scheme of Ref.~\cite{Coleman:1973jx}, in the region $\vev{S}\neq0$.
However, this anomalous dimension is the same as the anomalous dimension of the conformal field theory at $\vev{S}=0$ (and $m=0$).
If $m=0$, the full theory has a superconformal symmetry. In particular, the theory is invariant under the scaling transformation
\beq
S(x,\h) \to e^{\r \D_S}S(e^\r x,e^{\r/2}\h),
\eeq
and also similar transformation for other fields, where $\D_S$ is the scaling dimension of $S$, and $\r$ is a transformation parameter.
When the vev $\vev{S}$ is non-vanishing, the superconformal symmetry is {\it spontaneously} broken. However, the quantum effective action must have
the symmetry under the above transformation, as in the case of ordinary spontaneously broken global symmetries in field theory. 
Because the effective K\"ahler potential has the scaling dimension two, the effective K\"ahler potential respecting the scaling symmetry is given by
\beq
K_{\rm eff}^{\rm conf} \propto |S|^{\frac{2}{\D_S}}. \label{eq:kahlerbysymm}
\eeq
Comparing Eq.~(\ref{eq:effkahler}) and (\ref{eq:kahlerbysymm}), we obtain
\beq
\D_S=\frac{2}{2-{\tilde \g}_{S*}}=1+\frac{\g_{S*}}{2}.
\eeq
This relation between $\D_S$ and $\g_{S*}$ is completely the same as the relation in the conformal field theory at $\vev{S}=0$.

We give a more explicit perturbative computation
for the effective K\"ahler potential by taking a $\overline{\rm DR}$ scheme in \ref{app-A}, where the discussion in this subsection is confirmed at 1-loop level.
There, momentum dependences of the effective action are also investigated.

\subsection{Mass-dependent corrections and higher derivative terms}
The mass-dependent corrections may be written as
\beq
 K_{\rm eff}^{\rm mass} = |{\hat S}|^2
 f\left( |{{\hat m} / {\hat S}}| \right), \label{eq:masscorrection}
\eeq
where ${\hat S}$ and ${\hat m}$ denote such variables that are
independent of the renormalization point $M$, and are given by
\bea
{\hat S}&=&M^{{\tilde \g}_{S*}/2} S^{1-{\tilde \g}_{S*}/2}=(M^{\g_{S*}/2}S)^{1/(1+\g_{S*}/2)}, \\
{\hat m}&=&(M^{-\g_{P*}} m)^{1/(1-\g_{P*})}.
\eea
Note that we do not need the dynamical scale $\L$ in this dimensional argument,
since $\L$ is defined as
$\L^{3N_C-N_F-N'_F}=M^{3N_C-N_F-N'_F}\exp(-8\pi^2/g_*^2)$ and the gauge
coupling $g_*$ is constant.

In fact, the effective K\"ahler potential (\ref{eq:effkahler}) and the superpotential (\ref{eq:effsuper}) can also be rewritten by using $\hat{S}$ and $\hat{m}$.
It is easy to see that 
\beq
K_{\rm eff}^{\rm conf}=(1-{\tilde \g}_{S*}/2)^{-2}|\hat{S}|^2. \label{eq:effkahler'}
\eeq
The superpotential (\ref{eq:effsuper}) can be rewritten as follows. Using the relation between $\L$ and $M$, 
$\L^{3N_C-N_F-N'_F}=M^{3N_C-N_F-N'_F}\exp(-8\pi^2/g_*^2)$, we obtain
\beq
W_{\rm eff}=N_C \left[M^{3N_C-(1+\g_{S*}/2)N_F-(1-\g_{P*})N'_F}
\exp(-8\pi^2/g_*^2)(M^{-\g_{P*}}m)^{N'_F}(\l_* M^{\g_{S*}/2}S)^{N_F}\right]^{1/N_C}.
\eeq
Then using the relations $\g_{S*}+2\g_{Q*}=0$ and $3N_C-(1-\g_{Q*})N_F-(1-\g_{P*})N'_F=0$, which are the conditions for $\b$ functions of $\l$
and $g$ to vanish, we obtain
\beq
W_{\rm eff}=N_C\left[\exp(-8\pi^2/g_*^2){\hat m}^{(1-\g_{P*})N'_F}(\l_* {\hat S}^{(1+\g_{S*}/2)})^{N_F}\right]^{1/N_C}.\label{eq:effsuper'}
\eeq
In this form it is manifest that the dynamical superpotential is indeed renormalization group invariant~\cite{ArkaniHamed:1997ut}.

Because a mass-independent renormalization is used, effects of the
non-zero $m$ in the K\"ahler potential are all contained in the function $f\left( |{{\hat m}/{\hat S}}| \right)$,
which is non-singular in the limit $m\to 0$ in the case we have assumed above.
Then, the behavior of the effective K{\"a}hler potential
in the massless limit $m \rightarrow 0$
implies that the $K_{\rm eff}^{\rm mass}$
increases at most as $|{\hat S}|^2$ for $|S| \rightarrow \infty$. %
Hence it does not alter the leading behavior of the effective
K{\" a}hler potential from the conformal term
(\ref{eq:effkahler'})
in the regime of large $|S|$. We neglect the term (\ref{eq:masscorrection}) from now on.

Next let us consider the higher derivative terms in the K\"ahler potential. Because $S$ interacts only through massive quarks $Q,~\tilde{Q}$,
the effective K\"ahler potential may be expanded in powers of derivatives. Then, for example, terms like
\beq
\int d^4\h\frac{|D^2\hat{S}|^2}{|\hat{S}|^2} \sim \frac{|\hat{F}_S|^4}{|\hat{S}|^4}
\eeq
contribute to the potential of $S$, where $\hat{F}_S$ is the $F$ component of the superfield $\hat{S}$ and $D$ is the superderivative.
In fact, these terms are effectively suppressed by positive powers of $\hat{m}/\hat{S}$. To see this, let us first assume that these terms are negligible 
compared to the leading term coming from Eq~(\ref{eq:effkahler'}). Then, by Eq.~(\ref{eq:effsuper'}), the equation of motion of $\hat{F}_S$ gives
\beq
\hat{F}^*_S \sim {\hat m}^{(1-\g_{P*})\frac{N'_F}{N_C}} {\hat S}^{(1+\g_{S*}/2)\frac{N_F}{N_C}-1}={\hat S}^2\left(\frac{\hat m}{\hat S}\right)^{(1-\g_{P*})\frac{N'_F}{N_C}},
\eeq
where the relations $\g_{S*}+2\g_{Q*}=0$ and $3N_C-(1-\g_{Q*})N_F-(1-\g_{P*})N'_F=0$ have been used. Thus 
$|\hat{F}_S/\hat{S}^2|$ is suppressed by positive power of $|\hat{m}/\hat{S}|$, 
and our initial assumption of neglecting the higher order terms in $\hat{F}_S$ is justified.

\subsection{Value of the scaling dimension}

If all couplings are small, we can use perturbation to compute the anomalous dimension (or equivalently the scaling dimension) of $S$ as in \ref{app-A}.  
In fact, we do not assume weak couplings, because we allow large anomalous dimension of $S$ at the fixed point. Then the perturbative computations of
\ref{app-A} are not reliable.

Even then, we can determine $\D_S$ by using a general property of superconformal field theory. 
In $\mathcal{N}=1$ superconformal field theory, $U(1)_R$ charges  $R$ and scaling dimensions $\D$ of chiral (primary) operators are 
related by $\D=\frac{3}{2}R$.  
So we can obtain the scaling dimensions $\D$ from the $R$ charges of the chiral operators.

We list the $R$ charges of the fields in Table~\ref{table:1}.
We have imposed that the $R$ charges of $Q~(P)$ and $\tilde{Q}~(\tilde{P})$ are the same.
Even then, because of the existence of an axial $U(1)_A$ symmetry, the definition of $U(1)_R$ is ambiguous. We represent this ambiguity by
parametrizing the $U(1)_R$ by a parameter 
$x$ in Table~\ref{table:1}. But at the comformal fixed point, there is a unique $U(1)_R$ symmetry which appears
in  the superconformal algebra. This $U(1)_R$ is the one whose $R$ charges are related to conformal dimensions.

\begin{table}[Ht]
 \begin{center}
 \begin{tabular}{c|c|c|c|c}
 &$Q,~{\tilde Q}$&$P,~{\tilde P}$&$S$&$m$ \\ \hline
 $U(1)_R$&$1+x$&$1-(N_C+N_F x)/N'_F$&$-2x$&$2(N_C+N_F x)/N'_F$
 
 \end{tabular}
 \caption{$R$ charges of the fields. $x$ is a parameter which represents an ambiguity of the definition of $U(1)_R$. $x$ is completely fixed
if we require the $U(1)_R$ to be the one appearing in the superconformal algebra of the theory. The charge of $m$ seen as a spurion field is also listed.}
 \label{table:1} 
\end{center}
\end{table}

This $U(1)_R$ symmetry can be determined by the $a$-maximization technique~\cite{Intriligator:2003jj}. 
According to Intriligator and Wecht, we can obtain the value of $x$
for the superconformal $U(1)_R$ charges by (locally) maximizing the following combination of 't~Hooft anomalies:

\bea
a_{\rm trial} &\equiv& \frac{3}{32} (3\tr R^3-\tr R) \nonumber \\
& =& \frac{3}{32} [ (N_C^2-1)\{3-1\} + \{3(-2x-1)^3-(-2x-1)\}+2N_CN_F\{3x^3-x\} \nonumber \\[0.2cm]  
&&+2N_CN'_F\{ 3(-(N_C+N_F x)/N'_F)^3-(-(N_C+N_F x)/N'_F)\} ] .
\eea
After straightforward calculations, we obtain the result
\beq
x=\frac{2+N^2_CN^2_F/N'^2_F-\sqrt{(2+N^2_CN^2_F/N'^2_F)^2+(-4+N_CN_F-N_CN^3_F/N'^2_F)(8/9+N^3_CN_F/N'^2_F)}}{-4+N_CN_F-N_CN^3_F/N'^2_F}.
\label{eq:amaximization}
\eeq
From Table~\ref{table:1}, we can see that the $U(1)_R$ charge of $S$ is $-2x$, so the scaling dimension of $S$ is $\D_S=-3x$.
One can check that in the limit $N_C,N_F,N'_F \gg 1$ and $n \equiv 3N_C-N_F-N'_F=  \mathcal{O}(1)$, this exact expression for $\D_S=1+\g_{S*}/2$
coincides with Eq.~(\ref{eq:perturbativeg_S}). We list some numerical results in Table~\ref{table:2}.

\begin{table}[Ht]
\begin{center}
\begin{tabular}{c|c|c|c|c|c|c}
$(N_C,~N_F,~N'_F)$&(3,~2,~3)&(3,~2,~4)&(4,~3,~3)&(4,~3,~4)&(4,~3,~5)&(5,~3,~5)                \\ \hline 
$\D_S$&1.78&1.46&2&1.70&1.48&1.86
\end{tabular}
\caption{Scaling dimensions of $S$ for some values of $N_C$, $N_F$, and $N'_F$. In the text we have assumed that $N_C<N'_F$,
but we also list the scaling dimensions in the case of $N_C \geq N'_F$ here.}
\label{table:2}
\end{center}
\end{table}

In conformal field theories, there are unitarity bounds on scaling dimensions of operators~\cite{Mack:1975je} (see also Ref~\cite{Minwalla:1997ka}).
Scaling dimensions of gauge invariant operators must be equal to or greater than $1$. The relation $\D=\frac{3}{2}R$ 
determines the scaling dimensions of operators $P\tilde{P}$ and $P\tilde{Q}$ to be%
\footnote{Later we will introduce singlets $S^{\tilde j}_i$ and a tree level superpotential $W=\l S^{\tilde j}_{i}Q^i\tilde{Q}_{\tilde j}$.
Then the operators $Q^i\tilde{Q}_{\tilde j}$ are not chiral primary because of the equations of motion of $S^{\tilde j}_i$, 
$\frac{1}{4}\bar{D}^2(S^{\tilde j}_{i})^*=\l Q^i \tilde{Q}_{\tilde j}$.
Then $\D=\frac{3}{2}R$ cannot be used for this operator.
}
\beq
\D_{P\tilde{P}}=3[1-(N_C+N_F x)/N'_F],~~~~\D_{P\tilde{Q}}=\frac{3}{2}[(1+x)+1-(N_C+N_F x)/N'_F].
\eeq
By requiring $\D_{P\tilde{P}}\geq 1$ and $\D_{P\tilde{Q}}\geq 1$, we can obtain (assuming $N_F<N'_F$)
\beq
2-\frac{2(N_F+N'_F)-3N_C}{N_F} \leq -3x \leq 2 + \frac{2(N_F+N'_F)-3N_C}{N'_F-N_F}.
\eeq
This gives a bound on $\D_S=-3x$. This bound suggests that $N_C,~N_F$ and $N'_F$ must satisfy the relation
\beq
\frac{3}{2}N_C\leq N_F+N'_F,
\eeq
in order that there is a conformal fixed point in our theory, as in Ref.~\cite{Seiberg:1994pq}.

\subsection{Potential of $S$}

From the effective superpotential (\ref{eq:effsuper}) and the effective K\"ahler potential (\ref{eq:effkahler}), the potential of $S$ is given by
\bea
V(S)&=&\left( \frac{\q^2 K_{\rm eff}}{\q S \q S^*} \right)^{-1} \left|\frac{\q W_{\rm eff}}{\q S}\right|^2 \nonumber \\
&=& M^{-{\tilde \g}_{S*}} |N_Fm^{N'_F}\L^{3N_C-N_F-N'_F}\l^{N_F} |^{2/N_C} |S|^{{\tilde \g}_{S*}-2(1-N_F/N_C)}.
\eea 
Since ${\tilde \g}_{S*}$ is related to $\D_S$ by ${\tilde \g}_{S*}/2=(\D_S-1)/\D_S$, the potential is an increasing function of $S$ if the condition
\beq
\frac{\D_S-1}{\D_S} > 1-\frac{N_F}{N_C} \label{eq:condition}
\eeq
is satisfied. Using Eq.~(\ref{eq:amaximization}) (or Table~\ref{table:2}), one can check that there exist many sets of values $N_C,~N_F,~N'_F$ which satisfy 
this condition.

\section{Adding more singlets}

In the model of Ref.~\cite{Izawa:2009mj} and also in the present model, there are mesonic runaway directions $|\det Q^i\tilde{Q}_j|\to \infty$ with 
$S=0$ (and $\tr Q\tilde{Q}= Q^i\tilde{Q}_i=0$ by the equation of motion of $S$).
In Ref.~\cite{Izawa:2009mj}, there are two parameters, the gauge coupling $g$ and Yukawa coupling $\l$. 
By appropriately choosing these parameters, $S$ can be stabilized at local minima where the vev of $S$ is large enough so that 
the local minima can be parametrically stable. In the present case, there is essentially one parameter, the mass parameter $m$, 
because couplings are uniquely fixed by the conformal dynamics. 
The potential of $S$ is a monotonically increasing function of $S$ in the regime
where $m$ dependence in the K\"ahler potential can be neglected. 
Thus there is no local minimum in the regime where the vev of $S$ is large, and we have to worry about the mesonic runaway.

To avoid the above difficulty, we can introduce $N_F \times N_F$ singlets $S^{\tilde j}_{i}$ instead of one singlet $S$, and adopt a tree level superpotential
\beq
W=\l S^{\tilde j}_{i}Q^i \tilde{Q}_{\tilde j}+mP^a\tilde{P}_a . 
\eeq
Then, as in Ref.~\cite{Izawa:1996pk}, the vevs of $Q^i\tilde{Q}_{\tilde j}$ are blocked by the equations of motion of $S^{\tilde j}_{i}$, and there is no
danger of mesonic runaway.

Let us consider the low-energy effective
superpotential and  K\"ahler potential of this theory.
The superpotential is exactly determined to be
\beq
W_{\rm eff}=N_C \left(m^{N'_F} \L^{3N_C-N_F-N'_F}\det (\l S)\right)^{1 \o N_C}, \label{eq:effsuper2}
\eeq
where the $S$ denotes the matrix $(S^{\tilde j}_i)$.
This form yields a definite scaling behavior
$W_{\rm eff}(\r^{N_C} S)=\r^{N_F} W_{\rm eff}(S)$
for $\r>0$.
On the other hand, the effective K\"ahler potential may be much more
complicated than that in the case of one singlet. Nevertheless,
from the superconformal symmetry,
we see that the effective K\"ahler potential has a scaling behavior
$K_{\rm eff}(\r^{\D_S} S)=\r^2 K_{\rm eff}(S)$,
where we have neglected the corrections
due to the mass $m$ for our purposes, as is the case for one singlet.
Hence the K{\" a}hler metric is given by
\bea
g_{IJ^*}=\frac{\q^2 K_{\rm eff}}{\q S^I \q S^{J*}},
\eea
where the index $I$ collectively denotes the indices $i$ and ${\tilde j}$.

Let us investigate the potential of $S$:
\beq
V=\sum_{I,J} g^{I^*J}
\left(\frac{\q W_{\rm eff}}{\q S^I}\right)^*
\frac{\q W_{\rm eff}}{\q S^J},
\eeq
where $g^{I^*J}$ is the inverse of $g_{IJ^*}$.
It is straightforward to see that the ``vector''
${\q W_{\rm eff}}/{\q S^I}$ is nonzero for any direction $S$.%
\footnote{It is helpful that
we can diagonalize the fixed $S$
by means of the $SU(N_F)_Q \times SU(N_F)_{\tilde Q}$ symmetry.}
Furthermore, the superpotential forces the singlets $S^{\tilde j}_{i}$ to be a diagonal form $S^{\tilde j}_{i}\propto \d^{\tilde j}_{i}$,
up to $SU(N_F)_{Q}\times SU(N_F)_{\tilde Q}$ transformation, as in the model of Ref.~\cite{Izawa:1996pk}.
We assume that the K\"ahler potential is not so singular as to change this behavior. 
For example, if the effective K\"ahler potential is of the form%
\footnote{A 1-loop computation as in \ref{app-A} suggests that the effective K\"ahler potential is of the form (\ref{eq:ansatzkahler}).
However, the computation is not so reliable because there are many scales (vevs of $S^{\tilde j}_{i}$) unlike the case of one singlet.}
\beq
K_{\rm eff} \propto \tr\{(S^\dagger S)^{1/\D_S}\},\label{eq:ansatzkahler}
\eeq
one can check that the above assumption is indeed satisfied.
Therefore, in order to determine whether the runaway of singlets is stabilized,
it is enough to consider the overall scaling behavior of the potential,
\beq
V(\r S) = \r^{2\left({\D_S-1 \o \D_S}-1+{N_F \o N_C}\right)} V(S), 
\eeq
and see whether the $S$ runs away or not,
namely, the situation is almost the same as in the case of one singlet.

There is one important difference from the case of one singlet: the value of $\D_S$. 
The one-parameter family of $U(1)_R$ symmetries is the same as in
Table~\ref{table:1}, but $a_{\rm trial}$ and thus the superconformal $U(1)_R$ are 
different. Using $a$-maximization, the value of $x$ is obtained as
\beq
\frac{2N^2_F+N^2_CN^2_F/N'^2_F-\sqrt{(2N^2_F+N^2_CN^2_F/N'^2_F)^2+(-4N^2_F+N_CN_F-N_CN^3_F/N'^2_F)(\frac{8}{9}
N^2_F+N^3_CN_F/N'^2_F)}}{-4N^2_F+N_CN_F-N_CN^3_F/N'^2_F}. 
\eeq 
Several numerical results are listed in Table~\ref{table:3}.
\begin{table}[Ht]
\begin{center}
\begin{tabular}{c|c|c|c|c|c|c}
$(N_C,~N_F,~N'_F)$&(3,~2,~3)&(3,~2,~4)&(4,~3,~3)&(4,~3,~4)&(4,~3,~5)&(5,~3,~5)                \\ \hline 
$\D_S$&1.70&1.36&2&1.59&1.35&1.81
\end{tabular}
\caption{Scaling dimensions of $S$ for several values of $N_C$, $N_F$, and $N'_F$, in the theory with $N_F \times N_F$ singlets.}
\label{table:3}
\end{center}
\end{table}
Using these values of $\D_S$, we can check that there exist many sets of values of $N_C,~N_F,~N'_F$ which satisfy the condition (\ref{eq:condition}).

\section{Conclusions and discussion}

The runaway behavior of dynamical superpotential in certain gauge theories can be stabilized by quantum corrections to 
the K\"ahler potential, leading to dynamical supersymmetry breaking~\cite{Izawa:2009mj}. 
In this paper, we have extended such theories to the ones which have superconformal symmetry at a high-energy regime.
By doing this, the effective K\"ahler potential is well controlled even in the strong coupling theories, 
and we can obtain the condition, Eq.~(\ref{eq:condition}) (with $\D_S$ determined exactly), 
under which the stabilization by the K\"ahler potential occurs. Furthermore, there is no Landau pole problem of the Yukawa coupling, 
which the original model suffers from.

In the present model, the singlet $S$ has minima near the origin, $|{\hat S}| \lsim {\hat m}$. There the superconformal symmetry is explicitly broken and 
the theory is very strongly coupled. Thus the investigation of the theory near the minima is a rather difficult problem, although not completely impossible.
However, even without knowing anything about the minima, we can see that SUSY is broken, by the following Witten index~\cite{Witten:1982df} argument.   

Suppose that we add a term $\tr(\k S)=\k^i_{\tilde j}S^{\tilde j}_i$  to the superpotential of the theory.
Here $\k=(\k^i_{\tilde j})$ is some matrix with $\det \k \neq 0$. By taking $m$ and $\k$ to be very large, 
we can calculate the Witten index of this theory.
First, $m$ makes $P$ and $\tilde{P}$ to be very massive, so we can forget about them. Second, $\k$ gives vevs to $Q\tilde{Q}$, 
and if $N_F<N_C$, $Q,~\tilde{Q}$ and $S$ all become massive.\footnote{An easy way to see this is as follows. When $N_F<N_C$, 
we can describe the theory by using gauge invariant ``mesons'' $M=(M^i_{\tilde j})=(Q^i\tilde{Q}_{\tilde j})$, as long as $\det M\neq0$. 
Then, the superpotential $W_{\rm tree}=\l \tr(SM)+\tr(\k S)$ is nothing but the mass terms of $S$ and $M$, with $\vev{S}=0$ and $\vev{M}=-\l^{-1}\k$.
} 
The vevs of $Q\tilde{Q}$ break gauge symmetry $SU(N_C)$ to $SU(N_C-N_F)$.
Finally, at low energy we obtain a pure $SU(N_C-N_F)$ gauge theory. Thus the Witten index is $N_C-N_F$. We can explicitly see
$N_C-N_F$ vacua by adding to Eq.~(\ref{eq:effsuper2}) the term $\tr(\k S)$,
\beq
W_{\rm eff}=N_C (m^{N'_F} \L^{3N_C-N_F-N'_F}\det \l S)^{1 \o N_C} + \tr(\k S).
\eeq
By the equation of motion of $S$, we obtain
\beq
S=-\k^{-1} \left( \frac{\L^{3N_C-N_F-N'_F} m^{N'_F}}{\det(-\l^{-1}\k)} \right)^{\frac{1}{N_C-N_F}}. \label{eq:manyvacua}
\eeq 
Here the $N_C-N_F$ vacua are represented by the $(N_C-N_F){\rm th}$ root. 

Let us take a limit $\k \to 0$. Then, all the vacua of Eq.~(\ref{eq:manyvacua}) go to infinity. This does not necessarily mean that there is no SUSY vacuum
at finite vevs of the fields. But because the Witten index is $N_C-N_F$, and $N_C-N_F$ vacua go to infinity,
if SUSY vacua exist at finite vevs, there have to be the same number of ``bosonic vacua'' and ``fermionic vacua'', and 
these vacua may not be protected by any invariant of the theory, such as the Witten index.\footnote{Of course this argument is not rigorous, 
because the Witten index may not be well defined at $\k=0$.}
So, it seems unlikely that such vacua exist, and we can reasonably believe that SUSY is broken in our theory.

\section*{Acknowledgements}

I.K.-I. would like to thank T.~Kugo for valuable discussions.
This work was supported by the Grant-in-Aid for Yukawa International
Program for Quark-Hadron Sciences, the Grant-in-Aid
for the Global COE Program "The Next Generation of Physics,
Spun from Universality and Emergence", and
World Premier International Research Center Initiative
(WPI Initiative), MEXT, Japan.
The work of KY is supported 
in part by JSPS Research Fellowships for Young Scientists.

\appendix
\setcounter{equation}{0}
\renewcommand{\theequation}{\Alph{section}.\arabic{equation}}
\renewcommand{\thesection}{Appendix~\Alph{section}}
\section{Explicit perturbative computation}\label{app-A}

In this appendix, we explicitly compute the 1PI effective action of $S$ 
at 1-loop level using a $\overline{\rm DR}$ scheme \cite{Siegel:1979wq}. $\overline{\rm DR}$ is conventional in perturbative computations in SUSY field theories, 
and this scheme (like $\overline{\rm MS}$) is also known to be a simple example of mass-independent renormalization~\cite{Weinberg:1996kr}. 
It is assumed that the gauge and Yukawa coupling constants at the fixed point are small and perturbative calculation is reliable. 
The renormalization group (RG) argument done in this appendix is valid to all orders in perturbation theory.

First, consider the RG equations of the Yukawa coupling $\l$ and the gauge coupling $g$. We will establish that there is indeed
an infrared conformal fixed point.
By the perturbative non-renormalization theorem, the RG equation of $\l$ is
\bea
\b_{\l} &\equiv& M\frac{d }{d M }|\l|^2 \nonumber \\
&=& (\g_S+2\g_Q)|\l|^2, \label{eq:lbeta}
\eea
where $\g_S$ and $\g_Q$ are the anomalous dimensions of $S$ and $Q(\tilde{Q})$ respectively, and $M$ is a renormalization scale.
The RG equation of $g$ is given by the NSVZ $\b$ function~\cite{Novikov:1983uc} up to 2-loop order
\bea
\b_g & \equiv & M\frac{d }{d M }g^2  \nonumber \\
&=& -\frac{g^4}{8\pi^2} \frac{3N_C-(1-\g_Q)N_F-(1-\g_P)N'_F}{1-N_Cg^2/8\pi^2}, \label{eq:gbeta}
\eea
where $\g_P$ is the anomalous dimension of $P(\tilde{P})$.

The $\g_S,~\g_Q$, and $\g_P$ are given at 1-loop order by
\bea
\g_S = N_CN_F\frac{ |\l|^2}{8\pi^2},~~~~\g_Q=\frac{|\l|^2}{8\pi^2} - \frac{N_C^2-1}{N_C} \frac{g^2}{8\pi^2},~~~~\g_P= 
- \frac{N_C^2-1}{N_C} \frac{g^2}{8\pi^2}.
\eea
Using these values in Eqs.~(\ref{eq:lbeta},\ref{eq:gbeta}) and requiring that $\b_{\l}=\b_g=0$, we can obtain
\beq
\frac{|\l_*|^2}{8\pi^2} \simeq \frac{2n}{N_CN_FN_{\rm tot}},~~~~~\frac{g_*^2}{8\pi^2} \simeq \frac{n}{N_CN_{\rm tot}}. \label{eq:fixedcoup}
\eeq
Here $n \equiv 3N_C-N_F-N'_F$, $N_{\rm tot}\equiv N_F+N'_F$, and $\l_*$ and $g_*$ are the values of $\l$ and $g$ at the fixed point. 
We have assumed that $N_C,N_F,N'_F \gg 1$ and $n =  \mathcal{O}(1).$
One can easily check using Eqs.~(\ref{eq:lbeta},\ref{eq:gbeta})
that this fixed point is indeed infrared stable, that is, the couplings flow into it, not away from it.
The value of $\g_S$ at the fixed point is given by
\beq
\g_{S*} \simeq \frac{2n}{N_{\rm tot}}. \label{eq:perturbativeg_S}
\eeq

Next let us go to the calculation of the effective action of $S$.
At 1-loop level, only a $Q,~\tilde{Q}$ loop contributes to the 2-point function of $S$. One can easily compute the 2-point part of the effective K\"ahler potential
by first computing the 2-point 1PI diagram of $\vev{F_SF^*_S}$, the $F$ component of the chiral field $S$, and then infer the entire 
effective K\"ahler potential by using supersymmetry.
For completeness, we take the external momentum to be non zero. 
Using $\overline{\rm DR}$, the 1PI effective action $\G_{\rm 1-loop}$ is
obtained as%
\footnote{We follow the convention of Ref.~\cite{Wess:1992cp}.} 
\bea
\int d^4\h \int \frac{d^4p}{(2\pi)^4} \left[ 1- \frac{N_CN_F| \l |^2}{16 \pi^2} \int^1_0 dz \ln \left( \frac{m_Q^2+z(1-z)p^2}{M^2} \right) \right] 
|{\tilde S}(p,\h)|^2
+\cdots \label{eq:1loopeff}
\eea
where the background field $S(x,\h) = S_0 + {\tilde S}(x,\h)$
with $S_0$ as the zero mode of $S(x,\h)$,
\footnote{
We separate the background field $S$ to
$S_0$ and $\tilde S$ to make computation fairly explicit.}
$m_Q=| \l S_0 |$, 
and
\beq
{\tilde S}(p,\h)= \int d^4x  {\tilde S}(x,\h)e^{-ipx}.
\eeq 
The ellipsis in Eq.~(\ref{eq:1loopeff}) represents 
terms containing more than two $\tilde S$. 

We now consider the RG improvement of Eq.~(\ref{eq:1loopeff}). Suppose
that the RG improved form of Eq.~(\ref{eq:1loopeff}) is given by
\beq
\G=\int d^4\h \int \frac{d^4p}{(2\pi)^4} D(p,m_Q,M,\l,g) |{\tilde S(p,\h)}|^2+\cdots.  
\eeq
Then, taking into account the fact that $\G$ is RG invariant and $\tilde S$ is RG variant, with the anomalous dimension given by $\g_S$, 
the Callan-Symanzik (CS) equation for $D(p,m_Q,M,\l,g)$ is given by
\beq
\left(M \frac{\q}{\q M} + \g_Q m_Q \frac{\q}{\q m_Q} + \b_\l \frac{\q}{\q |\l|^2} + \b_g \frac{\q}{\q g^2} - \g_S \right)D(p,m_Q,M,\l,g) = 0, \label{eq:CSeq}
\eeq
where we have used the RG equation of $m_Q$, namely,
$M \frac{\q}{\q M} m_Q= \g_Q m_Q$. 

In the above analysis, we have shown that the theory flows into a conformal fixed point, which is infrared stable. 
Then the $\b$ functions $\b_\l$ and $\b_g$
vanish, and $\g_S$ and $\g_Q$ are constant.%
\footnote{The couplings might seem to run below the scale $m_Q$, because conformal
symmetry is spontaneously broken by the vev of $S$. In fact, $\overline{\rm DR}$ is a renormalization scheme
in which RG runnings of dimensionless couplings are not affected by mass terms, so that $\l$ and $g$ can really be constant.
}
In this case, Eq.~(\ref{eq:CSeq}) can be easily solved. If we rewrite $D(p,m_Q,M)$ as 
\beq
D(p,m_Q,M)=(M \hat{D} (p,\hat{m}_Q,M) )^{\g_S} ,
\eeq
where $\hat{m}_Q=(M^{-\g_Q}m_Q)^{1/(1-\g_Q)}$, and  
$\hat{D}(p,\hat{m}_Q,M)$ has mass dimension $-1$,
then Eq.~(\ref{eq:CSeq}) becomes
\beq
M \frac{\q}{\q M} \hat {D} (p,\hat{m}_Q,M) =0.
\eeq
Thus the solution of the CS equation is simply $\hat{D}=\hat{D}(p,\hat{m}_Q)$. This expression is valid to all orders in perturbation theory. 

Let us return to 1-loop computation. To obtain the 1-loop improved effective action, we have to obtain $\hat{D}(p,\hat{m}_Q)$ to zeroth order in small 
couplings.
Expanding the solution $D(p,m_Q,M)=(M \hat{D} (p,\hat{m}_Q) )^{\g_S}$ in powers of small couplings and comparing with 
Eq.~(\ref{eq:1loopeff}), we obtain
\beq
\g_S \ln (M \hat{D}(p,m_Q) ) = - \frac{N_CN_F| \l |^2}{16 \pi^2} \int^1_0 dz \ln \left( \frac{m_Q^2+z(1-z)p^2}{M^2} \right).
\eeq
From this equation, we can correctly obtain the 1-loop anomalous dimension $\g_S=N_CN_F|\l|^2/8\pi^2$, 
and our final expression for the RG improved $\G$ is
\bea
\G=\int d^4\h \int \frac{d^4p}{(2\pi)^4} \exp \left[ - \frac{\g_S}{2} \int^1_0 dz \ln \left( \frac{\hat{m}_Q^2+z(1-z)p^2}{M^2} \right) \right] 
|{\tilde S}(p,\h)|^2 + \cdots .\label{eq:improvedeff}
\eea

As a check, consider the limit $m_Q \to 0$. Then Eq.~(\ref{eq:improvedeff}) becomes
\bea
\G \to \int d^4\h \int \frac{d^4p}{(2\pi)^4} \exp(\g_S) \left( \frac{M^2}{p^2} \right)^{\g_S/2}
|{\tilde S}(p,\h)|^2 +\cdots .
\eea
From this effective action, we can compute the 2-point correlation functions. 
For example, the 2-point correlation function of the lowest component ${\tilde S}(p)$ of ${\tilde S}(p,\h)$ is
\beq
\vev{{\tilde S}(p) {\tilde S}^*(p')} = (2\pi)^4 \d^{(4)}(p-p') \exp(-\g_S) \frac{M^{-\g_S }}{p^{2-\g_S}}, 
\eeq
exactly as expected from conformal invariance of the theory at $S_0=0$.

Our real interest is the limit $S_0 \to \infty$. In this limit, Eq.~(\ref{eq:improvedeff}) becomes
\bea
\G&=&\int d^4\h \int \frac{d^4p}{(2\pi)^4} \left( \frac{M}{\hat{m}_Q} \right)^{\g_S}
|{\tilde S}(p,\h)|^2 +\cdots \nonumber \\
 &=&\int d^4\h \int d^4x \left( \frac{M}{|\l S_0|} \right)^{\g_S/(1+\g_S/2)}
|{\tilde S}(x,\h)|^2 +\cdots, \label{eq:result1}
\eea
where we have used the relation $\hat{m}_Q=(M^{-\g_Q}m_Q)^{1/(1-\g_Q)}$ and $\g_S+2\g_Q=0$ at the conformal fixed point. 
Eq.~(\ref{eq:result1}) is what we need to compute the potential of $S$ when the vev of $S$ is large.

In fact, if we assume that $\hat{D}(p,\hat{m}_Q)$ is non-singular in the limit $p \to 0$, which seems plausible because $S$ only interacts through massive
quarks $Q,~\tilde{Q}$, then dimensional analysis tell us that $\hat{D}(p=0,\hat{m}_Q) \propto \hat{m}_Q^{-1}$. Thus we can conclude that
\beq
\G =C\int d^4\h \int d^4x \left( \frac{M}{|\l S_0|} \right)^{\g_S/(1+\g_S/2)}
|{\tilde S}(x,\h)|^2 +\cdots, \label{eq:effkahlerapp1}
\eeq
to all orders in perturbation theory, where $C$ is some constant which may depend on coupling constants. From the 1-loop result, we know
$C=1+{\cal O}(g^2,|\l^2|)$.

The effective action should not depend on $S_0$ and ${\tilde S}(x,\h)$ separately, but depend only on the combination $S(x,\h)=S_0+{\tilde S}(x,\h)$. 
Thus the first term and the dots in Eq.~(\ref{eq:effkahlerapp1}) should combine together to give
\bea
\G={\tilde C} \int d^4\h \int d^4x M^{\g_S/(1+\g_S/2)}|S(x,\h)|^{2/(1+\g_S/2)}
+\textrm{higher-derivative terms},
\eea
where the higher-derivative terms come from expansion of
$\hat{D}(p,\hat{m}_Q)$ in powers of $p^2/\hat{m}_Q^2$, and $\tilde C$ is defined as
${\tilde C}=C(1+\g_S/2)^{-2}|\l|^{-\g_S/(1+\g_S/2)}$. This is the effective K\"ahler potential in the $\overline{\rm DR}$ scheme.

\end{document}